\documentclass[12pt,amsmath,amssymb]{revtex4}
\usepackage{graphicx} 
\usepackage{amsmath} 
\usepackage{color}
\begin{document}
\title{High temperature expansion, virial coefficients and exclusion statistics}
\author{ Saptarshi Mandal}
\address{ International Center for Condensed Matter Physics\\
Universidade de Brasilia, Caixa Postal 04667, 70910-900 Brasilia, DF, Brazil}
\begin{abstract}
 We follow the generalisation of exclusion statistics to infinite dimensional Hilbert space as envisaged in Phys. Rev. Lett. {\bf{72}}, 3629, 1994. We reproduce the third virial coefficients at leading order for single species of anionic gas and 2nd mixed virial coefficients for multicomponent anionic gas. We argue that this particular method  can be useful in determining definition of  mutual exclusion statistics. We  demonstrate this by taking high temperature expansion of two particle partition function of well known systems and show that it follows   Haldane's definition of exclusion statistics. We also discuss equilibrium particle distributions at thermodynamic limit.
\end{abstract} 
\maketitle
\section{Introduction}

How the elementary particles in nature and collective excitations in physical systems interact and what is the statistics obeyed by them has always been a subject of interest and active research. It has been known  that in 3 spatial dimension  the elementary particles  can only obey fermionic or bosonic statistics. In bosonic system a single particle quantum state can be occupied by arbitrary number of bosons, while no two identical fermions can occupy one and same quantum state by Pauli principle. But in recent years it has been recognized that particles with ``fractional statistics'' which is intermediate between bosons and fermions can exist in two-dimensional ~\cite{wilczek,khare} or in one-dimensional system ~\cite{yangyang,haldane1}. These exotic particles are named anyons whose statistics can be anything in between bosonic and fermionic statistics. The distribution function for N anyons is given by,
\begin{eqnarray}
W=\frac{(G+(N-1)(1-\alpha))!}{N!(G-\alpha N- (1-\alpha))!},
\label{ds1f}
\end{eqnarray}
with $\alpha=0$ corresponding to boson and $\alpha=1$ for fermions and for anyons $\alpha$ can take any value in between 0 and 1. Now if one has a mixture of several species of anyons then  mutual exclusion statistics may also happen due to non orthogonal localized states. In this situation single particle  available states of one species `i' can be changed  due to the presence of particles of another species `j'. F Haldane  ~\cite{haldane1} defined the statistical interactions for such systems by the linear relation,
\begin{eqnarray}
\Delta d_i&=& - \sum_j \alpha_{ij} \Delta N_j
\label{hd01}
\end{eqnarray} 
where $\Delta d_i$ is the reduction of the available single particle states for 'i'th species. $\Delta N_j$ is a set of allowed changes of the particle numbers of the species 'j'. $\alpha_{ij}$ is the mutual exclusion statistics parameter between 'i'th and 'j'th species. In general $\alpha_{ij} \ne \alpha_{ji}$.  The above definition allows us to write down the distribution function as follows,
\begin{equation}
\label{Dnh1}
D^{{[n_{i}]}}_{N}=\prod^{m}_{i}\frac{(d'_{1i}+(1-\alpha_{ii})(n_{i}-1))!}{n_{i}!(d'_{1i}-1-\alpha_{ii}(n_{i}-1))!}
\end{equation}
where  $d^{\prime}_{1i}$  is the effective dimension of the single particle Hilbert space for the 'i'th species and is given by, $d^{\prime}_{1i}=d_{1i} - \sum_{j\neq i}n_{j}\alpha_{ij}$, $d_{1i}$ is dimension of the one particle Hilbert space for `i'th species.  Here $[n_{i}]$ refers a particular distribution of different species of particle into total N particles such that $N= \sum n_i$.  Following this definition thermodynamics of anyonic gas and other properties have been investigated which can be found in ~\cite{yangshi,isakov}. In this article we would like to extend the procedure taken in ~\cite{shankar} where the definition of fractional exclusion statistics has been extended to a system with infinite dimensional Hilbert spaces.  In ~\cite{shankar} it has been shown that the high temperature expansion(HTE) of the partition function enables to  express the 2nd virial coefficients in terms of mutual exclusion statistical parameter. This method is quite novel in the sense that it gives us a first hand information about the dependency of partition function on the exclusion statistical parameters, without going into the details of calculation of partition function.  By a straight forward generalisation of the method shown in ~\cite{shankar}, we find that the mixed virial co-eficients is determined by the mutual exclusion statistics parameters $\alpha_{ij}$ and $\alpha_{ji}$. We also reproduce the third virial coefficients at leading  order.  Moreover we argue that this particular procedure can show how the mutual exclusion statistics works among different species of anyons. This is quite relevance as contrary to the original definition proposed by Haldane, it has been conjectured in ~\cite{victor1} that mutual exclusion statistical interaction works in a different way. According to ~\cite{victor1}, equation ~(\ref{hd01}) defining  mutual exclusion statistical interaction should be read as follows,
\begin{eqnarray}
\Delta d_i&=&- d_i \alpha_{ij} \Delta N_j,
\label{dvadf1}
\end{eqnarray}
which states the reduction of the single particle states of species `i' due  to the change of particle $\Delta N_j$ of species `j' is proportional to the dimension of single particles states of species `i'. We believe that the pedagogical procedure taken here can help to understand how mutual exclusion statistics works among different species of anyons. We follow the following outline in presenting this work.  In section 2, we will give a brief account of the main idea in  reference ~\cite{shankar} to establish a connection between the high temperature expansion of the partition function and the regulated definition of partition function obtained from the distribution function. We will generalise the above mentioned method to multispecies of anionic gases.  Further we show that it can be extended to higher  order and reproduce the third virial coefficients which matches well with the exact results obtained in ~\cite{isakov1} at leading order. In section 3, we discuss how one can determine the definition of exclusion statistics using this particular method.  We discuss the two particle partition function of two simple physical systems and show that this method reaffirms   the definition of mutual exclusion statistics as stated by Haldane ~\cite{haldane1}. On the basis of the results obtained in section 3 and 4,  we discuss  the equilibrium particle distribution
 in thermodynamic limit  obtained from the definition ~(\ref{hd01}) and ~(\ref{dvadf1}).  In section 5 we summarize our results.

\section{High temperature limit and virial coefficients}
Here we wish to find a correspondence between the high temperature expansion of the partition function and the distribution function in the thermodynamic limit. When the dimension of the Hilbert space is infinite then in the high temperature limit one can replace the distribution function by the corresponding partition function, as mentioned in {~\cite{shankar}}. Let us write down the distribution function for a single species for anyonic gas,
\begin{eqnarray}
D_{N}&=& \frac{(d +(1-\alpha)(N-1))!}{N!(d-1+\alpha(N-1))}
\label{eqd1}
\end{eqnarray}
In the above expression `d' represents the single particle dimension and `$\alpha$' is the exclusion statistical parameter. For Bose system one have $\alpha=0$ and for fermion we have $\alpha=1$. If one rewrite the equation (~\ref{eqd1}) in the descending power of $d^n$, one obtains,
\begin{eqnarray}
\frac{D_{N}}{d^N}&=&1+ \sum^{N}_{i=1}\sigma_{N,i} \frac{1}{d^{i}}
\label{dnaps1}
\end{eqnarray} 

Where one can easily find the coefficients  $\sigma_{N,i}$. They are in general polynomial of various order of `$\alpha$'.  One sees that the above equation contains
 information of the dependency on `$\alpha$' at various order of $1/d$. We wish to exploit this in the following. At first order it has been shown in ~\cite{shankar}, that it yields correct dependency of 2nd virial coefficients on `$\alpha$'. At the limit $d \rightarrow \infty$ we can write from ~(\ref{dnaps1}), at first order,
\begin{eqnarray}
\sigma_{0}= \frac{1}{2} -\alpha= lim_{d \rightarrow \infty} \frac{d}{N(N-1)} (N! \frac{D_{N(\alpha)}}{d^N}-1)
\label{fst1}
\end{eqnarray}
Now let us write the regulated definition of the Hilbert space by the corresponding N-particle partition function as we have,

\begin{eqnarray}
D_N= lim _{\beta \rightarrow 0} Z_{N}=lim_{\beta \rightarrow 0} Tr{e^{-\beta H_N}}
\label{gendef1}
\end{eqnarray}
where $\beta$ is the inverse temperature and $H_N$ denotes the N-particle Hamiltonian. Therefore we can write from ~(\ref{fst1}) and ~(\ref{gendef1}),
\begin{eqnarray}
\frac{1}{2}-\alpha= lim_{d \rightarrow \infty} \frac{C z_1}{N(N-1)} (N! \frac{Z_{N(\alpha)}}{z^N_1}-1)
\label{2ndeq1}
\end{eqnarray}
Here `C' is an overall constant which is found to be  $2^{d_s}$ where $d_s$  is spatial dimension. From now on we will focus on 2 dimension and work in Harmonic oscillator confinement.  Now  we see that r.h.s  of the above equation depends on  'N'. In ~\cite{shankar}, it has been shown that the r.h.s is actually related to case $N=2$. Keeping in mind that for $N=2$,  r.h.s reduces to 2nd virial coefficients $B_2= z_1(1- 2 \frac{Z_2}{z^2_1} )$ we get, following ~\cite{shankar},
\begin{eqnarray}
\frac{1}{2}-\alpha &=& -2 B_2
\end{eqnarray}

Now let us briefly discuss the multispecies genaralisation of this. It is easy to find that from the distribution function  ~(\ref{Dnh1}), we can write similar to equation ~(\ref{fst1}),
\begin{eqnarray}
-\sum_{i}n_{i}S_{i} +\sum_{i}n_{i}(n_{i}-1)(\frac{1}{2}-\alpha_{ii})=lim_{d \rightarrow \infty }d(\frac{D^{[n_{i}]}_{N}\Pi n_{i}!}{d^{N}}-1)
\label{mpsg1}
\end{eqnarray}
where $S_{i}=\sum_{i\neq j}n_{j} \alpha_{ij}$. Now using the same procedure to arrive at equation ~(\ref{2ndeq1}) from ~(\ref{fst1}),  and  using ~(\ref{fst1}) for each species we get(in Appendix we give a detail derivation of the following equation),
\begin{equation}
\label{gab4}
(g_{ab}+g_{ba})=-4Z_{1}(\frac{Z_{1_{a},1_{b}}}{Z^{2}_{1}}-1). 
\end{equation}
The r.h.s is nothing but four times the mixed 2nd virial coefficients ~\cite{isakov1}.  The above equation relates the mutual exclusion statistics parameter $\alpha_{ab}$ with the two particle partition function $Z_{1_{a},1_{b}}$. This has also been found earlier in ~\cite{isakov} in a different way. If '$a$' type and '$b$' type particles are statistically independent then $Z_{ab}=Z^{2}_{1}$ and then $\alpha_{ab} + \alpha_{ba}$ is zero which tells that $\alpha_{ab}$ and $\alpha_{ba}$ both should be zero. It could also be possible that $\alpha_{ab}=-\alpha_{ba}=p$, where $p$ is a constant, as referred in ~\cite{She}, and in  ~\cite{johnson}.  It is important to note that it is the combination $(\alpha_{ab}+\alpha_{ba})$ which appears together in Eq. ~(\ref{gab4}). However to get an unique expressions for individual parameter $\alpha_{ab}$ and $\alpha_{ba}$, we need to go to next order to calculate the third virial coefficients which is given in terms of 3 particle partition function and solve for $\alpha_{ab}$ and $\alpha_{ba}$.  To go the higher order we notice, keeping in mind $\sigma_0 \sim 1/2 -\alpha$, the high temperature expansion is equivalents to keeping only the first term of the following approximate equation for the N-particle partition function obtained from the distribution function ~(\ref{dnaps1}).
\begin{eqnarray}
\frac{Z_{N}}{{z^N}_1}&=& 1+ \sum^{N}_{i=1} C_i \sigma_{N,i}   \frac{1}{z^i_1}
\end{eqnarray} 
where $Z_N$ represents the `N' particle partition function and $z_1$ is the single particle partition function. $C_i$`s are the coefficients which is to be found to get the approximate expressions for $Z_N$ at various order `i'. At first approximation  we get $C_1=4$ and  neglect all other term in the r.h.s. Without going to the general derivation for $C_2$, we focus on just two particle and three particle partition function in the harmonic well potential. Then we get the value of $C_2=1/9$. The significance of number $9=3^2$, is that it denotes a states where 3 particles are in the same energy level. Then we can write the general expression for 2 particle and 3 particle partition function for a single species of anyonic,
\begin{eqnarray}
\frac{2Z_2}{z^2_1}&=& 1+ \frac{1}{4} \frac{1-2\alpha}{z_1} +  \frac{1}{9} \frac{\alpha(\alpha-1)}{z^2_1} \\
\frac{6Z_3}{z^3_1}&=&1 + \frac{1}{4} \frac{3(1-2\alpha)}{z_1} + \frac{1}{9} \frac{12\alpha(\alpha-1)+2}{z^2} + C_3  \frac{4\alpha(\alpha-1)(2\alpha-1)}{z^3_1}
\end{eqnarray}
Let us write down the expression for cluster coefficients ($\tilde{b_l}$) and virial coefficients $a_l$ following ~\cite{isakov1}, 
\begin{eqnarray}
\tilde{b_2}&=&\frac{1-2\alpha}{8 z_1} + \frac{1}{18 z^2_1}\alpha(\alpha-1)\\
\tilde{b_3}&=&\frac{1}{z^2_1}(\frac{1}{6}\alpha(\alpha-1) + \frac{1}{27})+\frac{C_3}{z^3_1}4\alpha(\alpha-1)(2\alpha-1)\\
a_3&=&\frac{1}{z^2_1} ((4^{-1-\delta} -2 .3 ^{-2-\delta})+(4^{-\delta} -3^{-\delta})\alpha(\alpha-1))\nonumber \\
&+&\rm{terms~with~coefficients~}~\frac{1}{z^3_1}~(\rm{ \delta~=1})
\end{eqnarray}

We see that at leading order  this  procedure reproduces the exact results obtained in ~\cite{isakov}. To get the higher order virial coefficients at leading order, we should keep track of all the sub leading terms of the lowest order virial coefficients. The sole purpose of the above exercise was to show that HTE of the partition function may be taken as a reliable method to extract the leading order dependency of the cluster and virial co-efficient as well. Also we see that the expansion of partition functions in terms of the single particle partition function contains important  information regarding how exclusion statistics works. At various order it depends on exclusion statistical parameters in a  specific way which can be utilized further. In the next section we will elaborate on this to show that it might help to understand how exclusion statistics works among different species of anyons. 

\section{HTE and two particle partition function}

In this section we wish to discuss about the definition of exclusion statistics and from the conclusion  of preceeding section we argue that HTE of partition function may be  a reliable method to understand it better. In a mixture of anyons of different species one can think of the effect of the exclusion statistics as a reduction of single particle dimension due to the presence of particle of other species. But how this single particle reduction happens is an important issue. F D Haldane ~\cite{haldane1} defined this reduction as given by ~(\ref{hd01}) which yields the distribution function ~(\ref{Dnh1}). We know that particles in a magnetic field in LLL obeys the definition ~(\ref{hd01}). For this reason we will present two particle partition function where each particle belongs to different anyonic species and show by HTE that indeed it follows defining equation ~(\ref{hd01}). Secondly we will take composite anions where each anion is attached with a different charge and flux tube ~\cite{arovas}. It has been argued that ~\cite{haldane1} such free anyons do not obey the distribution function ~(\ref{Dnh1}). However  when we confine such particles in a harmonic oscillator potential it indeed obeys the defining equation ~(\ref{hd01}) and ~(\ref{Dnh1}), at lowest order.  In ref ~\cite{victor1}, it has been argued that the definition ~(\ref{hd01}) and the distribution function which follows this definition leads to an ambiguous equlibrium particle distribution in thermodynamic limit which is surely not desirable. This has been resulted  to conjecture a new definition of mutual exclusion statistics given by ~(\ref{dvadf1}) which defers from the definition given in equation ~(\ref{hd01}).  This definition has also been analytically derived in ~\cite{victor2} in the context of multispecies generalisation of  Colagero-Sutherland Model(CSM) ~\cite{shankar1}. Now  the consequences of these two definition  will be manifested  in the distrubution function of two particle where each particle belongs to different species `1' and `2'. Let us write down the two different distribution function in two particle case as mentioned.
\begin{eqnarray}
D^h_{2}&=& \frac{(d_1- \alpha_{12})!}{(d_1-1-\alpha_{12})! } \frac{(d_2- \alpha_{21})!}{(d_2-1-\alpha_{21})!} \label{dsh2} \\
D^v_{2}&=& \frac{(d_1- d_1\alpha_{12})!}{(d_1-1-d_1\alpha_{12})! } \frac{(d_2- d_2\alpha_{21})!}{(d_2-1-d_2\alpha_{21})!} \label{dsv2}
\end{eqnarray} 
Here we have assumed that both the definition follows the same kind of combinatorial distribution ~\ref{Dnh1}. Now let us approximate equation ~(\ref{dsh2}) and ~(\ref{dsv2}) by   their corresponding partition function at appropriate limit following section 2. We get the following expressions at appropriate limit,
\begin{eqnarray}
Z^h_2& \sim& z^2_1 -c^h_1 (\alpha_{12} + \alpha_{21})z_1 + c^h_2 \alpha_{12} \alpha_{21} \label{hhh}\\  
Z^v_2&\sim & z^2_1(1-\alpha_{12})(1-\alpha_{21}) \label{vvv}
\end{eqnarray} 

Now we see clearly that two different definition may be compared by the  expression of the two particle partition function and their leading order dependencies on the single particle partition function. Now let us consider a physical system where fractional exclusion statistics can be realized. We  have already mentioned that particle in LLL under the magnetic field obeys fractional statistics. We take the 'symmetric' case $\alpha_{12}=\alpha_{21}$ and the case where  single particle dimension is same for each species without loss of generality. Two particle partition function in LLL is given by ~\cite{isakov1,ouvry},
  
\begin{eqnarray}
Z_2&=& e^{- \beta \alpha_{12} \tilde{w}} \frac{e^{-\beta w_t }}{1- e^{\beta \tilde{w}}} \frac{e^{-\beta w_t }}{1- e^{\beta \tilde{w}}}
\end{eqnarray}
Here $\tilde{w}= w_t- w_c,~~w_t=\sqrt{w^2+ w^2_c}$. 'w' is the harmonic oscillator frequency and $w_c$ is the cyclotron frequency. We assume $\tilde{w} \sim w$, so that $w_c << w$. Now we can expand the right hand side in the limit $\beta \rightarrow 0$ and expand in the power of $z_1 \sim \frac{1}{\beta w}$ to get the following expression,
\begin{eqnarray}
Z_2&\sim& z^2_1 -c^h_1 \alpha_{12}z_1 + c^h_2 \alpha^2_{12}
\end{eqnarray}

We see that it follows closely the definition given by Haldane.  Now we will   show another example as described below. The model for a describing  anyons follows  ~\cite{arovas}. One imagines that each anyon is a composite object of point charge and flux in the units of `e' and `$\phi_0$', where `e' and $\phi_0$ denotes the electronic charge and the magnetic flux quanta. Then a particular species of anyon is described by the amount of charge $\alpha e$ and $\beta \phi_0$ it carries. Different anyonic species are distinguished by the pair $\alpha$ and $\beta$. They can be anything in general. In the symmetric case which demands $\alpha_1 \beta_2 = \alpha_2 \beta_1$ one can easily solve for the two particle Hamiltonian in harmonic oscillator confinement. In this case the   Hamiltonian gets separated into center of mass system and a relative particle coordinate system. Then one finds the following expressions for the  center of mass partition function,
\begin{equation}
Z_{cm}=\frac{\cosh(\frac{\beta w}{2})}{2\sinh(\beta w) \sinh(\frac{\beta w}{2})}.
\end{equation} 
The relative particle partition function  is given by,
\begin{equation}
 Z_{rel}=\frac{\cosh(\frac{\beta w(1-\alpha_{12})}{2})}{2\sinh(\beta w) \sinh(\frac{\beta w}{2})}.
\end{equation} 
In calculating the partition function we have taken into account all values of '$l$' even and odd, so $\alpha_{12}$ should vary from '0' to '2'. Now the two particle partition function $Z_2=Z_{cm} Z_{rel}$ , at the high temperature limits turns out to be,
\begin{eqnarray}
\frac{Z_{2}}{z^2_1}& = & 1 + \frac{(1-\alpha_{12})^2}{2 z_1} + \rm{~higher ~order~ terms}\frac{1}{z^2_1} 
\label{2pp}
\end{eqnarray}

Now we comparing with equation  ~(\ref{hhh}) and ~(\ref{vvv}), we find it has a close connection with equation ~(\ref{hhh}). The difference may be due to the reason that we have not worked in the LLL in the presence of magnetic field. This might suggests many important aspect to think of. It might quite be possible that the particular procedure taken here is not sufficient or the distribution function is quite different for the exclusion statistical defined by equation ~\ref{dvadf1}. Also it is pertinent to mention that here we are concerning about 2 particle  case which is not thermodynamic limit. But then question arises whether exclusion statistics works in a different manner when going from few particle case to many particle case in thermodynamic limit. It may also happen  that conjecture in ref ~\cite{victor1} is satisfied by certain class of anionic species but not the example taken here. In that case it is a very important findings and may indicate that exclusion statistics works in a different way than what we are used to think. 
\section{On the equilibrium particle distribution in the thermodynamic limit}

Now we wish to discuss the equilibrium particle distribution obtained from the two definition ~(\ref{hd01}) and ~(\ref{dvadf1}). From the definition ~(\ref{hd01}) and the distribution function ~(\ref{Dnh1}) we get the  following coupled equations for the equilibrium particle distributions ~\cite{isakov}.
\begin{eqnarray}
e^{\frac{\epsilon_k - \mu_k}{kT}}&=& (1+ w_k) \prod_i (\frac{w_i}{1+w_i})^{\alpha_{ik}} \label{dsth1}\\
n_k&=&\frac{1}{w_k + \alpha_{kk}} (1- \frac{1}{G_k} \sum_{i \ne k} G_i \alpha_{ki} n_i) \label{dsth2}
\end{eqnarray}
Here $n_i= N_i /G_i$, is the equilibrium particle density for the 'i'th species. $N_i$ is the total number of particle that belong to the same species and $G_i$ is the corresponding dimension of the single particle Hilbert space. We need to solve these coupled equation for every 'i'. If we look at the second equation we see a dependency on the dimensions of the Hilbert space which implies that if one changes the dimension of one species a redistribution of equilibrium particle density happens among different species which is not desirable.  Now let us write down the equilibrium particle distribution from the definition of ~(\ref{dvadf1}), which is given by ~\cite{victor1},
\begin{eqnarray}
\beta(\mu_k -\epsilon_k) + ln \frac{(1+n_k)^{1-\alpha_{kk}}}{n_k}&=& \sum_{j(\ne k)} G_{j} ln(1+n_{j}) \alpha_{jk} \label{dstv1}
\end{eqnarray}

Still here we find a dependency of the equilibrium population number on the dimension of the Hilbert space of each species. But the equation ~(\ref{dstv1}) can be converted to an appropriate integrals. In ref ~\cite{victor1} it has been argued that the equation ~(\ref{dsth1}) and ~(\ref{dsth2}) leads to an ambiguous determination of equilibrium population number. In that study a system of two species of anion has been considered. Let $N_1$ and $N_2$ are the number of anions belonging to the species `1' and `2'.  Then the species of `2' has been conceived as a union of two groups lets say $N_{2a}$ and $N_{2b}$ such that $N_2=N_{2a}+N_{2b}$. The original  system  is thought of a system of three  species with mutual exclusion statistical parameter between every two  species. Then if one compares the original set of equations with the new set of equations it leads to inconsistencies as far as equilibrium population density is concerned. One can recall that the above procedure is a reminiscent of the so called `Gibbs Paradox` ~\cite{patheria}. In the spirit of `Gibbs Paradox' we can think of dividing the original system into two subsystem with particle $\frac{N_1 + N_2}{2}$ in each subsystem  separated from each other. This can be achieved by a  completely  `reversible' process and thus thermodynamically giving same limit. Then it is easy to check that the system of equation ~(\ref{dsth1}) and ~(\ref{dsth1}) does not lead inconsistencies. This is due to the following reason. The equilibrium density $n_1= N_1/G_1$ remains unchanged also the ratio $G_1 /G_2$ remains unchanged for each subsystem and in the full system after each subsystem are adiabatically merged. But if we apply this simple idea to equation ~(\ref{dstv1}), we find that in equation ~(\ref{dstv1}) $n_j$ remains same for each subsystem but in the r.h.s. of equation ~(\ref{dstv1}), we get $G_j/2$ instead of $G_j$ which is different from the parent equation. We can not get back the original equation from this two daughter equation.  The main objective is to emphasis that it is a very important issue as mentioned in ~\cite{victor1} and should get due attention. If there is any two anyonic species  which follows the definition ~(\ref{dvadf1}), the two particle partition function should reveal more information when appropriately expanded in terms of the single particle partition function. Now taking lessons from ~\cite{victor1} we wish to consider a 'hypothetical' situation  which interpolates these two definitions. Let us  assume that there are a class of anyons which obey the following definition given by,
\begin{eqnarray}
\Delta d_i&=& - \alpha_{ij} \frac{d_i}{d_j}  N_j
\label{defsap1}
\end{eqnarray}
Where `$d_j$' is the single particle dimension of the species `j' and $\alpha_{ij}$ is the mutual exclusion statistics parameter between the species 'i' and 'j'. The equilibrium particle distribution obtained from it is given by,
\begin{eqnarray}
e^{\frac{\epsilon_k - \mu_k}{kT}}&=& (1+ w_k) \prod_i (\frac{w_i}{1+w_i})^{\frac{G_i \alpha_{ik}}{G_k}} \label{sapd1} \\
n_{k}&=& \frac{1}{w_k + \alpha_{kk}} (1- \sum_{i \ne k} \alpha_{ki} n_i) \label{sapd2}
\end{eqnarray}
We need to solve this coupled equation for every species `i'.
Now still we have a volume dependency on the equilibrium particle distribution as $w_k$ is determined by the equation ~(\ref{sapd1}). But the effect of it is to renormalize the $n_k$ in a trivial manner as can be seen from the above equation  ~(\ref{sapd2}). The equilibrium particle distribution does not undergo much redistribution if the dimension of single particle states is changed. We wish to mention that  for the  moment it is just a speculation. The above system of equation reduces to those of ~(\ref{dsth1}) and ~(\ref{dsth2}) for a single component anionic gas thus recovering well known distribution function for Bose and Fermi system. Besides this trivial limit it might be important to consider the following situation.
Let us imagine that in addition to the  condition $N_i <<G_j$ for every pair we have one anyonic species 'k' for which $G_k >> G_i,~~i\ne k$. Then for the species 'k', the effect of
exclusion statistics would be minimal and its equilibrium particle population should be closely related to its original particle population before mixing with other species of anyons. Now from equation ~(\ref{sapd1}), we find that in the r.h.s we can neglect all other term except the term i=k , and we have,
\begin{eqnarray}
e^{\frac{\epsilon_k- \mu_k}{kT}}(1+w_k)^{\alpha_{kk}-1} = w^{\alpha_{kk}}_{k}
\label{sapd3}
\end{eqnarray}

The above equation is nothing but the equation for the equilibrium population for the 'k'th species only. Then from equation ~(\ref{sapd2}), we clearly see that $n_k$ is  determined by its original distribution multiplied by  a modulating  factor $(1- \sum_{i \ne k} \alpha_{ki} n_i)$. It is also to be noted that if we consider a two particle system each belonging to different species with Hilbert dimension $d_1$ and $d_2$ respectively, we get the following distribution for a symmetric matrix $\alpha_{12}=\alpha_{21}$.
\begin{eqnarray}
W_{12}&=& (d_1- \alpha_{12} \frac{d_1}{d_2}) (d_2- \alpha_{12} \frac{d_2}{d_1}) \\
&=& (d_1- \alpha_{12}) (d_2- \alpha_{12})
\end{eqnarray}
In the above equation  2nd line is nothing but the distribution function obtained from Haldane's definition. It remains to see whether there is any physical system which obeys the mutual fractional statistics defined by equation ~(\ref{defsap1}). 
\section{Discussion}
In the present work we have followed generalisations of the definition of exclusion statistics to infinite dimensional Hilbert as first conceived in ~\cite{shankar}. We have reproduced the earlier results ~\cite{isakov1} that mixed 2nd virial coefficients is determined by the mutual exclusion statistics parameter. We reproduced third virial coefficients at leading order. We have shown that this particular method can help us understanding mutual exclusion statistics works by concentrating on two particle partition function. We have reestablished the well known results of particles in LLL in  magnetic field and anions attached with charge and flux tube in a harmonic oscillator potential. Though we have worked with system confined in a harmonic well potential, we believe the procedure taken here may be extended in other physical system and two particle partition function contains the information how mutual  exclusion statistics works among anions. We have shown that the mutual exclusion statistics follows the definition by F Haldane ~\cite{haldane1}, in contrary to what has been conjectured in ~\cite{victor1}. However it might quite be possible that the particular example taken in ~\cite{victor2} in view of the conjecture in ~\cite{victor1} is an exception or the particular method taken here to establish the definition of mutual statistics is not sufficient.  Later we show that the dependency of the equilibrium population resulting from  the definition followed in ~\cite{haldane1} and ~\cite{victor1} can be further minimized by a hypothetical definition given in equation ~(\ref{defsap1}). Though we have not found any realization of this definition it would be interesting to see if there are any.  

\vspace{4cm}
\begin{center}
\appendix{{\bf{Appendix}}}
\end{center}
\vspace{1cm}
Here we give a detail derivation of equation ~(\ref{gab4}).
Taking the limit $\beta\rightarrow 0 $ we get from  Eq.~(\ref{mpsg1}),
\begin{equation}
\label{Zn}
-\sum_{i}n_{i}S_{i} +\sum_{i}n_{i}(n_{i}-1)(1/2-g_{ii})=\lim_{\beta\rightarrow 0}CZ_{1}(\frac{\Pi n_{i}! Z_{[n_{i}]}}{Z_{1}^{N}}-1), 
\end{equation}
where $Z_{1}$ is the single particle partition function($Z_{1}=\frac{e^{-\beta w}}{(1-e^{\beta w})^{2}}$) and $Z_{[n_{i}]}$ is the 'N' particle partition function for a given distribution
$[n_{i}]$ among the `m' different species. The  constant 'C' is an overall constant of proportionality. The value of 'C' is 4 for a system confined in harmonic potential in two dimension ~\cite{shankar}. The following high temperature expansion of $N$ particle partition function in Harmonic  oscillator confinement  has been used in ~\cite{shankar},
\begin{equation}
\label{Zn2}
\frac{Z_{N}}{Z^{N}_{1}}=\frac{1}{N!}+f^{N}_{2} (\beta w)^{2} +f^{N}_{3}(\beta w)^{4}+....
\end{equation}
Here $Z_{N}$ is the $N$ particle partition function. Generalising  this for multispecies system, we write the  following high temperature expansion for the factor $Z_{[n_{i}]}$,
\begin{equation}
\label{Zn1}
\frac{Z_{[n_{i}]}}{Z^{N}_{1}}=\frac{1}{(\Pi^{m}_{i}n_{i}!)}+f^{[n_{i}]}_{2} (\beta w)^{2} +f^{[n_{i}]}_{3}(\beta w)^{4}
\end{equation}
where coefficients $f^{[n_{i}]}_{k}$ are to be determined and here we consider the expansions in powers of $(\beta w)^{2}$
as we will be considering the particles to be confined in an oscillator potential in two space dimensions. Now if we sum the
above expressions for different  set of distribution set $[n_{i}]$ we get
\begin{equation}
\label{Zn2}
\frac{Z_{N}}{(Z^{N}_{1})}=\frac{m^{N}}{N!}+ (F^{N}_{2}(\beta w)^{2}++F^{N}_{3}(\beta w)^{4}+......
\end{equation}
where $F^{N}_{2}=\sum_{[n_{i}]} f^{[n_{i}]}_{2}$ and $Z_{N}=\sum_{[n_{i}]}Z^{[n_{i}]}_{N}$.
Now Eq.~(\ref{Zn}),~(\ref{Zn1}), and ~(\ref{Zn2}) gives,
\begin{equation}
\label{gij1}
 -\sum_{i\neq j}\alpha_{ij} +\sum_{i}(\frac{1}{2}-\alpha_{ii})= CZ_{1}(\beta w)^{2}F^{N}_{2} \frac{(N-2)!}{m^{(N-2)}}.
\end{equation}
For our purpose to determine the mutual exclusion statistics parameter $g_{ij}$ it is sufficient to consider a system having two species,$i$ and $j$. From now on we would call these species as `$a$' and `$b$'. Then the above equation would be,
\begin{equation}
\label{gab1}
-(g_{ab}+g_{ba})+(\frac{1}{2}-g_{aa})+(\frac{1}{2}-g_{bb})=4Z_{1}(\beta w)^{2} F^{N}_{2}\frac{(N-2)!}{2^{N-2}}.
\end{equation}
Which reduces to ,$\frac{1}{2}-g=4 Z_{1}(\beta w)^{2} f^{N}_{2}\frac{(N-2)!}{2^{N-2}}$, for a single component system as it should be. This $  f^{N}_{2}$  would relate the $ g$ to the second virial coefficient. Now we will show that $F^N_2$ is actually related to $F^2_2$.

 The grand canonical partition function of this two component system is given by,
\begin{equation}
\label{grand}
Z=\sum^{\infty}_{n_{a}=0}\sum^{\infty}_{n_{b}=0}e^{-\beta(\mu_{a}n_{a}+\mu_{b}n_{b})}Z_{n_{a},n_{b}}
=\sum^{\infty}_{n_{a}=0}\sum^{\infty}_{n_{b}=0}z^{n_{a}}_{a} z^{n_{b}}_{b} Z_{n_{a},n_{b}},
\end{equation}
 where  $z_{a}=e^{-\beta\mu_{a}}$, $z_{b}=e^{-\beta\mu_{b}}$. $\mu_{a}$ and $\mu_{b}$ are the chemical potential of the `a' type and `b' type of anion  respectively. $z_{a}$ and $z_{b}$ are the respective fugacity parameter. The equation of the state of the system in the fugacity expansion may be written as ~\cite{patheria}
\begin{equation}
\label{fugacity}
\beta P=\frac{1}{V} ln(Z)=(\frac{Z_{1}}{V})\sum^{\infty}_{l_{a}+l_{b}=1}b_{l_{a},l_{b}}z^{l_{a}}_{a}z^{l_{b}}_{b}.
\end{equation}
 The general expression for $b_{l_{a},l_{b}}$ is given by
\begin{equation}
b_{l_{a},l_{b}}=(Z^{l-1}_{1})\sum_{p_{i}}(-1)^{(\sum_{i_{a}}p_{i}-1)}(\sum_{i_{a}}p_{i}-1)!\Pi_{i}
(Z_{i_{a},i_{b}}/Z^{i_{a}+i_{b}}_{1})^{p_{i}}/p_{i}!.
\end{equation}
The summation over $p_{i}$ is constrained by $\sum^{l_{a}}_{i_{a}=0}i_{a}p_{i}=l_{a}$ and
$\sum^{l_{b}}_{i_{b}=0}i_{b}p_{i}=l_{b}$ with an additional condition $l_{a}+l_{b}\ge 1$. Now substituting the expression for $Z_{i_{a},i_{b}}/Z^{i_{a}+i_{b}}_{1}$ in the above equation we get the following expressions for $b_{l_{a},l_{b}}$,
\begin{equation}
\label{blab}
b_{l_{a},l_{b}}=\frac{1}{(\beta w)^{2l-2}} [\sum^{l_{a}+l_{b}-2}_{n_{a}+n_{b}=0}(-1)^{n_{a}+{n_{b}}} 
\frac{f^{l_a-n_a,l_{b}-n_{b}}_{2}}{n_{a}!n_{b}!} (\beta w)^{2}+....]
\end{equation}
If we demand all $b_{l_{a},l_{b}}$ to be  remain finite the term up to the power $(\beta w)^{2l-2}$ should go to zero, so in particular the coefficient of $(\beta w)^{2}$ should be zero. This leads to the following relations
\begin{equation}
\label{f2}
f^{n_{a},n_{b}}_{2} = \frac{n_{a}n_{b}}{n_{a}!n_{b}!} f^{1_{a},1_{b}}_{2}+\frac{n_{a}(n_{a}-1)}{n_{a}!n_{b}!} f^{2_{a},0}_{2}+\frac{n_{b}(n_{b}-1)}{n_{a}!n_{b}!} f^{0,2_{b}}_{2}
\end{equation}
The above equations when summed up gives 
\begin{equation}
\label{F2}
F^{N}_{2}=\frac{2^{N-2}}{(N-2)!}(f^{0,2_{b}}_{2}+f^{2_{a},0}_{2}+f^{1_{a},1_{b}}_{2})=\frac{2^{N-2}}{(N-2)!}F^2_2
\end{equation}
 So from Eq.~(\ref{Zn1}) and Eq.~(\ref{gab1}) we get,
\begin{eqnarray}
\label{gab3}
-(g_{ab}+g_{ba})+(\frac{1}{2}-g_{aa})+(\frac{1}{2}-g_{bb})= 4Z_{1}(\frac{Z_{1_{a},1_{b}}}{Z^{2}_{1}}-1) \nonumber \\
+ 4 Z_{1}(\frac{Z_{2_{a},0}}{Z^{2}_{1}} -\frac{1}{2})+4 Z_{1}(\frac{Z_{0,2_{b}}}{Z^{2}_{1}} -\frac{1}{2})
\end{eqnarray}

Now consideration of one component system ex. only of $a$ species, would give us,
\begin{equation}
(\frac{1}{2}-g_{aa})=4Z_{1}(\frac{Z_{2a}}{Z^{2}_{1}}-\frac{1}{2}).
\end{equation}
 Similarly for $b$ species we have,
\begin{equation}
(\frac{1}{2}-g_{bb})=4Z_{1}(\frac{Z_{2b}}{Z^{2}_{1}}-\frac{1}{2}).
\end{equation}
Using the above relation we get the following relation for the mutual exclusion statistical parameter,
\begin{equation}
\label{gab41}
(g_{ab}+g_{ba})=-4Z_{1}(\frac{Z_{1_{a},1_{b}}}{Z^{2}_{1}}-1). 
\end{equation}
\end{document}